\documentstyle[12pt]{article}
\advance\hoffset by -7mm
\makeatletter
\@addtoreset{equation}{section}
\makeatother
\renewcommand{\theequation}{\thesection.\arabic{equation}}
\renewcommand{\title}[1]{\null\vspace{25mm}

\noindent{\Large{\bf #1}}\vspace{10mm}

\noindent {\large By }}
\newcommand{\authors}[1]{\noindent{\large #1}\vspace{20mm}

}
\newcommand{\address}[1]{\noindent #1\vspace{5mm}

}
\renewcommand{\abstract}[1]{\vspace{17mm}

\noindent{\small{\em Abstract.} #1}\vspace{2mm}

}

\newcommand{\journal}[4]{{\em #1~}{\bf #2}\,(19#3)\,#4.}

\newcommand{\ijmp}{\journal {Int. J. Mod. Phys.}}

\newcommand{\cmp}{\journal {Comm. Math. Phys.}}

\newcommand{\np}{\journal {Nucl. Phys.}}
\newcommand{\pl}{\journal {Phys. Lett.}}
\newcommand{\mpl}{\journal {Mod. Phys. Lett.}}
\newcommand{\prep}{\journal {Phys. Reports}}

\newcommand{\annp}{\journal {Ann. Phys. (N.Y.)}}

\makeatletter
\@addtoreset{equation}{section}
\makeatother
\renewcommand{\theequation}{\thesection.\arabic{equation}}

\newcommand{\lp}{\left(}\newcommand{\rp}{\right)}
\newcommand{\lc}{\left[}\newcommand{\rc}{\right]}
\newcommand{\lac}{\left\{}\newcommand{\rac}{\right\}}

\newcommand{\D}{\Delta}
\renewcommand{\a}{\alpha}
\renewcommand{\b}{\beta}
\renewcommand{\d}{\delta}
\newcommand{\e}{\varepsilon}

\newcommand{\m}{\mu}
\newcommand{\n}{\nu}

\renewcommand{\o}{\omega} \renewcommand{\O}{\Omega}
\newcommand{\p}{\psi}
\newcommand{\r}{\rho}
\newcommand{\s}{\sigma} \renewcommand{\S}{\Sigma}
\renewcommand{\t}{\theta}

\renewcommand{\AA}{{\cal A}}
\newcommand{\BB}{{\cal B}}

\newcommand{\FF}{{\cal F}}
\newcommand{\GG}{{\cal G}}
\newcommand{\HH}{{\cal H}}

\newcommand{\NN}{{\cal N}}

\newcommand{\PP}{{\cal P}}

\newcommand{\RR}{{\cal R}}
\newcommand{\SS}{{\cal S}}

\newcommand{\WW}{{\cal W}}

\newcommand{\complex}{{\kern .1em {\raise .47ex
\hbox {$\scriptscriptstyle |$}}
    \kern -.4em {\rm C}}}
\newcommand{\real}{{{\rm I} \kern -.19em {\rm R}}}
\newcommand{\rational}{{\kern .1em {\raise .47ex
\hbox{$\scripscriptstyle |$}}
    \kern -.35em {\rm Q}}}

\newcommand{\cb}{{\bar c}}

\newcommand{\dfrac}[2]{{\displaystyle{\frac{#1}{#2}}}}

\newcommand{\tfrac}[2]{{\textstyle{\frac{#1}{#2}}}}

\newcommand{\sla}{\raise.15ex\hbox{$/$}\kern -.57em}

\newcommand{\twiddle}{\lower.9ex\rlap{$\kern -.1em\scriptstyle\sim$}}

\newcommand{\vf}{{\varphi}}

\newcommand{\equ}[1]{(\ref{#1})}

\newcommand{\eq}{\begin{equation}}
\newcommand{\eqn}[1]{\label{#1}\end{equation}}
\newcommand{\eea}{\end{eqnarray}}
\newcommand{\eqa}{\begin{eqnarray}}
\newcommand{\eqan}[1]{\label{#1}\end{eqnarray}}
\newcommand{\ba}{\begin{array}}
\newcommand{\ea}{\end{array}}
\newcommand{\eqac}{\begin{equation}\begin{array}{rcl}}
\newcommand{\eqacn}[1]{\end{array}\label{#1}\end{equation}}

\def\non{\nonumber\\}
\def\BS{\BB_{\S}}
\def\ST{\widetilde{\S}}
\def\vfb{\bar{\vf}}
\def\cb{\bar{c}}
\def\6{\partial}
\def\={\!\!\!&=&\!\!\!}
\def\+{\!\!\!&&\!\!\!+~}
\def\xx{\!\!\!&&\!\!\!-~}

\begin{document}

\setcounter{page}{0}
\thispagestyle{empty}\hspace*{\fill} REF. TUW 94-10

\hspace*{\fill} UGVA-DPT 1994/07-858
\title{A Renormalized Supersymmetry in the Topological Yang-Mills
       Field Theory}
\authors{A. Brandhuber$^\#$, O. Moritsch$^{\ddag}$\footnote{Work
supported
         in part by the
         ``Fonds zur F\"orderung der Wissenschaftlichen Forschung''
         under Contract Grant Number P9116-PHY.},
         M.W. de Oliveira$^{\ddag}$\footnote{Work supported
         in part by the ``Fonds zur F\"orderung der
Wissenschaftlichen
         Forschung'', M085-Lise Meitner Fellowship.},
         O. Piguet$^{\dag}$\footnote{Work supported in part by the
Swiss
          National Science Foundation.} \\ and M. Schweda$^{\ddag}$}
\address{$^\#$ CERN, CH-1211 Gen{\`e}ve 23 (Switzerland)}
\address{$^{\dag}$ D{\'e}partement de Physique Th{\'e}orique,
         Universit{\'e} de Gen{\`e}ve\\
         24, quai Ernest Ansermet, CH-1211 Gen{\`e}ve 4
(Switzerland)}
\address{$^{\ddag}$ Institut f\"ur Theoretische Physik,
         Technische Universit\"at Wien\\
         Wiedner Hauptstra\ss e 8-10, A-1040 Wien (Austria)}
\abstract{We reconsider the algebraic BRS
          renormalization of Witten's topological
          Yang-Mills field theory by making
          use of a vector supersymmetry Ward identity which improves
          the finiteness properties of the model.
          The vector supersymmetric structure is a common feature of
          several topological
          theories. The most general local counterterm is determined
          and is shown to be a trivial BRS-coboundary.}


\newpage
\section{Introduction}

The introduction of Witten's topological Yang-Mills field theories
in the recent past~\cite{wit}, provided another
interesting example
of the fruitful interplay between mathematics and physics.
With regard to mathematics, these models were devised in order to
give a field theoretical interpretation for the Donaldson invariants
of four-manifolds~\cite{don}.
{}From the point of view of physics, topological field theories
could,
in principle, give rise to a new class of gravity lagrangians in
which
the metric tensor would merely play the role of a gauge parameter.
Such a remarkable property would essentially be an indication of
unbroken
general covariance at the quantum level~\cite{wit,bbrt}.

Mainly for this last reason, the renormalization aspects of
Witten's
topological model have been widely investigated by several groups
in the last few years~\cite{bms,hor,brt,bbt,mad}.
In ref.~\cite{wer}, one of the authors
discussed the ultraviolet behaviour of the theory in the framework of
the algebraic BRS technique~\cite{brs,bec,pr}, which is known to be
regularization independent. In particular, one was able to determine,
to all orders, the most general local counterterm to the classical
action
in a Landau type gauge. It was then proven the absence of anomalies
and
that the cohomological nature of the model was totally insensitive
to quantum effects.

In the present paper, we will be concerned in extending the analysis
of ref.~\cite{wer}, by exploiting an additional invariance of the
theory:
namely topological vector supersymmetry which together with
BRS-symmetry and translations build a supersymmetry algebra of
Wess-Zumino type~\cite{br,dgs}.
This supersymmetric structure, whose
topological origin is manifest when a Landau gauge is chosen, is a
common feature of a large class of topological
models~\cite{dgs,ms,ssw}.
One will realize that the associated supersymmetric Ward identity
significantly improves, with respect to~\cite{wer}, the finiteness
properties
of the theory, restraining the number of monomials in the fields
which show
up in the counterterm expression. It is worthwhile to underline once
more
that our approach does not make reference to any regularization
scheme and extends to all orders of perturbation theory.

The work is organized as follows: in Section 2 we describe the
particular Landau gauge-fixing in the classical approximation
and its quantization, in Section 3 we show the absence of anomalies
in the theory and we obtain the local counterterm. Section 4 contains
some concluding remarks. We also devote two appendices to a better
understanding of the counterterm construction.

\section{The Classical Action and the Landau Gauge}

Let us start by presenting the BRS framework of Witten's topological
Yang-Mills field theory in $D=4$ Euclidean space. We adopt here the
standard construction of refs.~\cite{bs,lp,osb} and introduce the
following set of fields: a gauge connection $A^a_\m$, an
anticommuting vector field $\psi^a_\m$ (also called topological
ghost),
a pair of antiself-dual tensor fields $(B^a_{\m\n},\chi^a_{\m\n})$
and two
ghost fields $(c^a,\vf^a)$. One needs also a couple of Lagrange
multipliers $(b^a,\eta^a)$ and a couple of
antighosts $(\cb^a,\vfb^a)$.
These variables will take values in the adjoint representation of an
arbitrary compact gauge group $\bf{G}$ with structure constants
$f^{abc}$.
We give their respective dimensions and Faddeev-Popov
ghost charges $(\Phi\Pi)$ in Table 1:

\begin{table}[h]
\begin{center}
\begin{tabular}{|c|c|c|c|c|c|c|c|c|c|c|}\hline
&$A_\m$&$\psi_\m$&$c$&$\vf$&$\chi_{\m\n}$&$B_{\m\n}$
&$\bar{\vf}$&$\eta$&$\cb$&$b$ \\ \hline
dimension&1&1&0&0&2&2&2&2&2&2 \\ \hline
$\Phi\Pi$&0&1&1&2&$-1$&0&$-2$&$-1$&$-1$&0 \\ \hline
\end{tabular} \\
\mbox{} \\
Table 1: Dimensions and Faddeev-Popov ghost charges of the fields.
\end{center}
\end{table}

The nilpotent BRS transformations for the fields defining the model
were originally obtained by Baulieu and Singer~\cite{bs}
(see also~\cite{lp}) and read as:

\eq
\begin{array}{ll}
{}~&~\\
sA^a_\m=-(D_\m c)^a+\psi^a_\m \ ,  &  \\
{}~&~\\
s\psi^a_\m=f^{abc}c^b\psi^c_\m+(D_\m \vf)^a \ ,  &  \\
{}~&~\\
s\vf^a=f^{abc}c^b\vf^c \ ,   &  \\
{}~&~\\
s c^a=\tfrac{1}{2}f^{abc}c^b c^c +\vf^a \ ,   &   \\
{}~&~\\
s\chi^a_{\m\n}=B^a_{\m\n} \ ,  &  sB^a_{\m\n}=0 \ , \\
{}~&~\\
s\bar{\vf}^a =\eta^a \ ,     &   s\eta^a=0 \ ,  \\
{}~&~\\
s\cb^{a}=b^{a} \ ,      &   sb^a=0 \ . \\
{}~&~
\end{array}
\eqn{BRS_TRANSFORMATIONS}

As one can also infer from ref.~\cite{bs}, in Euclidean flat space
Witten's classical action takes the form of a pure gauge fixing term
which enforces the topological properties of the (anti-)instantonic
field
configurations of $A^a_\m$.
In the present work we propose the following Landau type gauge-fixed
action:
\eqa
S_{\rm gf}\=s\int d^{4}x\lc \chi^{a\m\n}F^{+a}_{\m\n}
+\bar{\vf}^a(\6_\m\psi^\m)^a+\cb^a\6_\m A^{a\m}\rc= \non
\=\int d^{4}x \lc B^{a\m\n}F^{+a}_{\m\n}
-\chi^{a\m\n}(\d^\r_\m\d^\s_\n
+\tfrac{1}{2}\e_{\m\n}^{~~\r\s})(D_\r\psi_\s)^a
+\chi^{a\m\n}f^{abc}F^{+b}_{\m\n}c^c \right. \non
&&~~~~~~+~\eta^a\6^\m\psi^a_\m
+\bar{\vf}^a\6^\m (D_\m \vf)^a
+\bar{\vf}^a \6^\m (f^{abc} c^b \psi_\m^c)
+b^a\6^\m A^a_\m\frac{}{} \non
&&~~~~~~+\left.\cb^a\6^\m(D_\m c)^a-\cb^a\6^\m\psi^a_\m \rc \ ,
\eqan{S_GF}
where
\eq
F^{+a}_{\m\n}=\tfrac{1}{2}(F^{a}_{\m\n}+\widetilde{F}^{a}_{\m\n}) \
\eqn{F_ANTISELF_DUAL}
stands for the antiself-dual part of the Yang-Mills field strength
with
\eq
\widetilde{F}^{a}_{\m\n}=\tfrac{1}{2}\e_{\m\n}^{~~\r\s}F^{a}_{\r\s} \
{}.
\eqn{F_TILDE}

It should be noticed however that $S_{\rm gf}$ displays a significant
difference with respect to the original and well-known action for the
topological gauge field theory~\cite{bs,lp,bms,wer}:
the gauge choice imposed on the
topological ghost makes use of the ordinary space-time partial
derivative,
viz.,
\eq
\6_\m\psi^{a\m}=0 \ ,
\eqn{TOP_GAUGE}
instead of the gauge-covariant one.
One will see shortly that the main motivation for such a modification
relies on the fact that the action \equ{S_GF} possesses a larger
content
of symmetries, in particular supersymmetry, than that of
the covariant situation~\cite{wer}.
We also stress
that condition \equ{TOP_GAUGE} is as acceptable as any other choice
of gauge
condition since a different one should not modify the physical
output. Indeed, in this work one is
mainly concerned with a better understanding of the ultraviolet
behaviour of
the model. The possible consequences to the structure of the
topological
observables~\cite{wit} that may eventually be produced by
\equ{TOP_GAUGE}
are not analyzed here.

In order to translate the BRS invariance of the gauge-fixed
action \equ{S_GF} into a
Slavnov identity, one follows a general rule~\cite{bec}, coupling the
non-linear parts of the transformations \equ{BRS_TRANSFORMATIONS} to
a set
of external sources $(\O^{a\m},\tau^{a\m},L^a,D^a)$. Their dimensions
and
Faddeev-Popov charges are defined in Table 2:

\begin{table}[h]
\begin{center}
\begin{tabular}{|c|c|c|c|c|}\hline
            & $\O^\m$ & $\tau^\m$ &  $L$  &  $D$  \\ \hline
 dimension  &    3    &    3      &   4   &   4   \\ \hline
 $\Phi\Pi$  &  $-1$   &  $-2$     & $-2$  & $-3$   \\ \hline
\end{tabular} \\
\mbox{} \\
Table 2: Dimensions and Faddeev-Popov ghost charges of the
external sources.
\end{center}
\end{table}

The invariant external action is then chosen to be:
\eqa
S_{\rm ext}\=\int d^{4}x\lc\O^{a\m}(D_\m c)^a+\tau^{a\m}(s\psi_\m^a)
+\tfrac{1}{2}L^a f^{abc} c^b c^c + D^a(s\vf^a)\rc \ ,
\eqan{S_EXT}
where one imposes a BRS-doublet structure for the transformation laws
of the sources,
\eq
\begin{array}{ll}
{}~&~\\
s\tau_\m^a=\O_\m^a \ ,~~~    &   s\O_\m^a=0 \ , \\
{}~&~\\
sD^a=L^a \ ,   &   sL^a=0  \ . \\
{}~&~
\end{array}
\eqn{BRS_SOURCES}
The complete action is then given by an exact BRS-variation:
\eqa
\S\=S_{\rm gf}+S_{\rm ext}= \non
\=s\int d^{4}x
\lc\chi^{a\m\n}F_{\m\n}^{+a}+\bar{\vf}^a\6_\m\psi^{a\m}
+\cb^a\6_\m A^{a\m}+\tau^{a\m}(D_\m c)^a
+\tfrac{1}{2}D^a f^{abc} c^b c^c\rc \ ,
\eqan{ACTION}
and satisfies the Slavnov identity:
\eq
\SS(\S)=0 \ ,
\eqn{SLAVNOV1}
with,
\eqa
\SS(\S)\=\int d^{4}x\lp\psi^a_\m\frac{\d\S}{\d A^a_\m}
-\frac{\d\S}{\d \O^{a\m}}\frac{\d\S}{\d A^a_\m}
+\vf^a\frac{\d\S}{\d c^a}+\frac{\d\S}{\d L^a}\frac{\d\S}{\d c^a}
+\frac{\d\S}{\d \tau^{a\m}}\frac{\d\S}{\d \psi^a_\m}
+\frac{\d\S}{\d D^a}\frac{\d\S}{\d \vf^a}\right. \non
&&~~~~~~+\tfrac{1}{2}B^a_{\m\n}\frac{\d\S}{\d \chi^a_{\m\n}}
+\eta^a\frac{\d\S}{\d \bar{\vf}^a}+b^a\frac{\d\S}{\d \cb^a}
+\left.\O^{a}_\m\frac{\d\S}{\d \tau^a_\m}+L^a\frac{\d\S}{\d D^a}\rp \
{}.
\eqan{SLAVNOV2}
{}From the Slavnov identity one reads off the linearized BRS
operator:
\eqa
\BS\=\int d^{4}x\lp\psi^a_\m\frac{\d}{\d A^a_\m}
-\frac{\d\S}{\d \O^{a\m}}\frac{\d}{\d A^a_\m}
-\frac{\d\S}{\d A^a_\m}\frac{\d}{\d \O^{a\m}}
+\vf^a\frac{\d}{\d c^a}+\frac{\d\S}{\d L^a}\frac{\d}{\d c^a}
+\frac{\d\S}{\d c^a}\frac{\d}{\d L^a}\right. \non
&&~~~~~~+\frac{\d\S}{\d \tau^{a\m}}\frac{\d}{\d \psi^a_\m}
+\frac{\d\S}{\d \psi^a_\m}\frac{\d}{\d \tau^{a\m}}
+\frac{\d\S}{\d D^a}\frac{\d}{\d \vf^a}
+\frac{\d\S}{\d \vf^a}\frac{\d}{\d D^a}
+\tfrac{1}{2}B^a_{\m\n}\frac{\d}{\d \chi^a_{\m\n}} \non
&&~~~~~~+\left.\eta^a\frac{\d}{\d \bar{\vf}^a}+b^a\frac{\d}{\d \cb^a}
+\O^{a}_\m\frac{\d}{\d \tau^a_\m}+L^a\frac{\d}{\d D^a}\rp \ .
\eqan{LINEAR_BRS}
Let us remark that it is a nilpotent operator:
\eq
\BS\BS=0 \ .
\eqn{BS_NILPOTENCY}
The Landau gauge-fixing conditions are:
\eq
\frac{\d\S}{\d b^a}=\6^\m A^a_\m \ ,
\eqn{GF_1}
and
\eq
\frac{\d\S}{\d \eta^a}=\6^\m \psi^a_\m \ .
\eqn{GF_2}
As usual~\cite{pr}, by (anti)commuting the gauge
conditions \equ{GF_1}-\equ{GF_2} with the
Slavnov identity \equ{SLAVNOV1}, one gets two ghost equations:
\eq
\frac{\d\S}{\d \cb^a}-\6^\m\frac{\d\S}{\d \O^{a\m}}=-\6^\m \psi^a_\m
\ ,
\eqn{GHOST_EQU_1}
and
\eq
\frac{\d\S}{\d \bar{\vf}^a}-\6^\m \frac{\d\S}{\d \tau^{a\m}}=0 \ .
\eqn{GHOST_EQU_2}

The complete action $\S$ obeys, in addition, a set of global
constraints.
They are derived by making use of the algebraic structure of the
Landau gauge-fixings imposed on the fields $A^a_\m$ and $\psi^a_\m$.
The first one is an antighost equation~\cite{bps}, which controls the
coupling of the ghost $\vf$:
\eq
\GG^a\S=\D^a_{\GG}
\eqn{ANTIGHOST_1}
where,
\eq
\GG^a=\int d^4 x \lp \frac{\d}{\d\vf^a}
-f^{abc}\bar{\vf}^b \frac{\d}{\d b^c} \rp
\eqn{G}
and
\eq
\D^a_{\GG}=\int d^4 x f^{abc}\lp \tau^{b\m}A^c_\m+D^b c^c \rp \ .
\eqn{DELTA_G}
We remark that $\D^a_{\GG}$ is a classical breaking, for it is linear
in
the quantum fields. This means that it can be promoted to
the full quantum theory without specific renormalization.

Commuting the antighost equation \equ{ANTIGHOST_1} with
the Slavnov identity \equ{SLAVNOV1}, one
obtains a further condition on $\S$ (again classically broken):
\eq
\FF^a\S=\D^a_{\FF},
\eqn{F_CONDITION}
with
\eqa
\FF^a\=\int d^4 x\lp\frac{\d}{\d c^a}-f^{abc}\bar{\vf}^b\frac{\d}{\d
\cb^c}
-f^{abc}A_\m^b\frac{\d}{\d \psi^c_\m}
-f^{abc}\tau^b_\m\frac{\d}{\d \O^c_\m}\right. \non
&&~~~~~~-\left.f^{abc}c^b\frac{\d}{\d \vf^c}-f^{abc}D^b\frac{\d}{\d
L^c}
+f^{abc}\eta^b\frac{\d}{\d b^c}\rp
\eqan{F}
and
\eq
\D^a_{\FF}=\int d^4 x f^{abc}\lp D^b\vf^c-\O^{b\m}A^c_\m
-\tau^{b\m}\psi^c_\m-L^b c^c \rp \ .
\eqn{DELTA_F}

Moreover, there exists a second antighost equation, controlling the
coupling of the ghost $c^a$:
\eq
\GG^{\prime a}\S=\D^a_{\GG^\prime}
\eqn{ANTIGHOST_2}
where,
\eq
\GG^{\prime a}=\int d^4 x \lp \frac{\d}{\d c^a}
+\tfrac{1}{2}f^{abc}\chi^{b\m\n}\frac{\d}{\d B^{c\m\n}}
+f^{abc}\bar{\vf}^b \frac{\d}{\d \eta^c}
+f^{abc}\cb^b\frac{\d}{\d b^c} \rp
\eqn{G_PRIME}
and
\eq
\D^a_{\GG^{\prime}}=\int d^4 x f^{abc}\lp D^b\vf^c-\O^{b\m}A^c_\m
-\tau^{b\m}\psi^c_\m-L^b c^c \rp \ .
\eqn{DELTA_G_PRIME}

By anticommuting \equ{ANTIGHOST_2} with \equ{SLAVNOV1} one sees
that $\S$ is left invariant under the rigid gauge symmetry of the
theory:
\eq
\RR^a\S=0 ,
\eqn{RIGID}
with
\eq
\RR^a=\sum_{\rm all~fields~\Phi}\int d^4 x
f^{abc}\Phi^b\frac{\d}{\d\Phi^c} \ .
\eqn{R}

\newpage
As announced in the introduction, the complete action $\S$ exhibits,
besides the BRS invariance \equ{BRS_TRANSFORMATIONS}, an ulterior
global
symmetry, the topological vector supersymmetry, whose generators
$\d_\a$
bear a Lorentz index. Its action on the fields and sources
is:
\eq
\begin{array}{ll}
{}~&~\\
\d_\a A^a_\m = 0 \ ,~~~~~~~ &   \d_\a \psi^a_\m = \6_\a A^a_\m \ ,
\\
{}~&~\\
\d_\a c^a = 0 \ ,              &   \d_\a \vf^a = \6_\a c^a \ ,  \\
{}~&~\\
\d_\a \chi^a_{\m\n} = 0 \ , & \d_\a B^a_{\m\n}=\6_\a \chi^a_{\m\n} \
, \\
{}~&~\\
\d_\a \bar{\vf}^a = 0 \ ,   &   \d_\a \eta^a = \6_\a \bar{\vf}^a \ ,
\\
{}~&~\\
\d_\a\cb^a=\6_\a\bar{\vf}^a \ , & \d_\a b^a=\6_\a\cb^a-\6_\a\eta^a \
, \\
{}~&~\\
\d_\a \tau^a_\m = 0 \ ,    &   \d_\a \O^a_\m = \6_\a \tau^a_\m \ ,
\\
{}~&~\\
\d_\a D^a = 0 \ ,       &   \d_\a L^a = \6_\a D^a  \ . \\
{}~&~
\end{array}
\eqn{VECTOR_SUSY}
The generators $\d_\a$ satisfy, together with the
BRS operator, a supersymmetry algebra:
\eq
\{s,\d_\a\}=\6_\a \ .
\eqn{DECOMPOSITION}
Let us note, that this algebra is automatically valid off-shell.
The associated supersymmetry Ward identity writes
\eq
\WW_\a\S=0 ,
\eqn{SUSY_WARD}
where
\eqa
\WW_\a\=\int d^4 x \lc (\6_\a A^a_\m)\frac{\d}{\d\psi^a_\m}
+(\6_\a c^a)\frac{\d}{\d\vf^a}
+\tfrac{1}{2}(\6_\a \chi^a_{\m\n})\frac{\d}{\d B^a_{\m\n}}
+(\6_\a \bar{\vf}^a)\frac{\d}{\d\eta^a}\right. \non
&&~~~~~~+\left. (\6_\a \cb^a -\6_\a \eta^a)\frac{\d}{\d b^a}
+(\6_\a \bar{\vf}^a)\frac{\d}{\d\cb^a}
+(\6_\a \tau^a_\m)\frac{\d}{\d\O^a_\m}
+(\6_\a D^a)\frac{\d}{\d L^a}\rc \ .
\eqan{W}

\noindent
We summarize the above results: the complete action $\S$ in
\equ{ACTION} fulfills

\noindent
\begin{tabular}{ll}
(i)     &  the Slavnov identity
\end{tabular}
\eq
\SS(\S)=0 \ ,
\eqn{SLAVNOV3}
\begin{tabular}{ll}
(ii)    &  the two gauge-fixing conditions,
           eqs.\equ{GF_1}-\equ{GF_2}, \\
 ~&~\\
(iii)   &  the two ghost equations,
           eqs.\equ{GHOST_EQU_1}-\equ{GHOST_EQU_2}, \\
 ~&~\\
(iv)   &  the first antighost equation
\end{tabular}
\eq
\GG^a\S=\D^a_{\GG} \ ,
\eqn{ANTIGHOST_3}
\begin{tabular}{ll}
(v)    &  the $\FF$-Ward identity
\end{tabular}
\eq
\FF^a\S=\D^a_{\FF} \ ,
\eqn{F_CONDITION_3}
\begin{tabular}{ll}
(vi)     &  the second antighost equation
\end{tabular}
\eq
\GG^{\prime a}\S=\D^a_{\GG^\prime} \ ,
\eqn{ANTIGHOST_4}
\begin{tabular}{ll}
(vii)    &  the rigid invariance
\end{tabular}
\eq
\RR^a\S=0
\eqn{RIGID_3}
and,

\noindent
\begin{tabular}{ll}
(viii)   &  the global supersymmetry Ward identity
\end{tabular}
\eq
\WW_\a\S=0 \ .
\eqn{SUSY_WARD_3}

We close this section by displaying the graded Lie algebra obeyed by
the
functional operators $\BS$, $\GG^a$, $\FF^a$, $\GG^{\prime a}$,
$\RR^a$
and $\WW_\a$:

\eq
\begin{array}{lll}
{}~&~&~\\
                      &  \lac \BS,\BS \rac = 0 \ ,   &  \\
{}~&~&~\\
\lc \GG^a,\BS \rc = \FF^a \ ,  &~&  \lac \FF^a,\BS \rac = 0 \ ,  \\
{}~&~&~\\
                      &  \lc \GG^a,\GG^b \rc = 0 \ ,   &  \\
{}~&~&~\\
\lc \GG^a,\FF^b \rc = 0 \ ,  &~&  \lac \FF^a,\FF^b \rac = 0 \ ,  \\
{}~&~&~\\
                &  \lac \GG^{\prime a},\GG^{\prime b} \rac = 0 \ ,
&  \\
{}~&~&~\\
\lac \GG^{\prime a},\BS \rac = \RR^a \ ,  &~&  \lc \RR^a,\BS \rc = 0
\ ,  \\
{}~&~&~\\
          & \lc \RR^a,\GG^{\prime b} \rc = -f^{abc}\GG^{\prime c} \ ,
& \\
{}~&~&~\\
\lc \RR^a,\RR^b \rc=-f^{abc}\RR^c \ ,  &~&  \lc\GG^a,\GG^{\prime
b}\rc=0 \ ,  \\
{}~&~&~\\
          & \lac \FF^a,\GG^{\prime b} \rac = -f^{abc}\GG^c \ , & \\
{}~&~&~\\
\lc \RR^a,\GG^b\rc=-f^{abc}\GG^c \ ,  &~&
\lc\RR^a,\FF^b\rc=-f^{abc}\FF^c \ ,
 \\
{}~&~&~\\
          & \lac \WW_\a,\BS \rac = \PP_\a \ , & \\
{}~&~&~\\
\lac \WW_\a,\WW_\b \rac = 0 \ ,  &~&  \lc \GG^a,\WW_\a \rc = 0 \ ,
\\
{}~&~&~\\
          & \lac \FF^a,\WW_\a \rac = 0 \ , & \\
{}~&~&~\\
\lac \GG^{\prime a},\WW_\a \rac = 0 \ ,  &~&  \lc \RR^a,\WW_\a \rc =
0 \ ,  \\
{}~&~&~
\end{array}
\eqn{ALGEBRA}
where $\PP_\a$ denotes the Ward operator for translations in the
space
of fields,
\eq
\PP_\a=\sum_{\rm all~fields~\Phi}\int d^4 x
(\6_\a\Phi)\frac{\d}{\d\Phi} \ .
\eqn{P}

\section{BRS Cohomology: Anomalies and Counterterms}

The purpose of this section is to present a systematic procedure for
the evaluation of the possible local counterterms to the complete
action
$\S$ in \equ{ACTION}.
One has to observe that the structure of any quantum
correction will be entirely governed by the set of classical
constraints
on $\S$, as long as they can be extended to the quantum level.
This latter statement implies that the absence of Slavnov and Ward
identity anomalies has to be demonstrated previously.

With this aim, we adopt the general recipe~\cite{bbbcd}
of collecting all the symmetry operators derived in the last section
into a unique functional operator which is nilpotent by construction.
It turns out then that the discussion of the renormalization
properties
of the model is significantly simplified when this strategy is
implemented.
The first step consists of the introduction of three pairs of global
ghosts $(\xi^{\m}, \t^{\m})$, $(u^{a}, v^{a})$ and $(w^{a}, y^{a})$
with
the following dimensions and $\Phi\Pi$ assignments:
\begin{table}[h]
\begin{center}
\begin{tabular}{|c|c|c|c|c|c|c|}\hline
           & $\xi$ &  $\t$ &  u   &   v   &   w   &  y  \\  \hline
 dimension & $-1$  & $-1$  &  0   &   0   &   0   &  0  \\  \hline
$\Phi\Pi$  &   2   &   1   &  3   &   2   &   2   &  1  \\  \hline
\end{tabular} \\
\mbox{} \\
Table 3: Dimensions and Faddeev-Popov charges of the
global ghosts.
\end{center}
\end{table}

With the help of these ghosts one builds up the
operator $\d$ as below:
\eqa
\d\=\BS+\xi^{\a}\WW_{\a}+\t^{\a}\PP_{\a}+u^{a}\GG^{a}+v^{a}\FF^{a}
+w^{a}\GG^{\prime a}+y^{a}\RR^{a}-\xi^{\m}\frac{\6}{\6 \t^{\m}} \non
\xx u^{a}\frac{\6}{\6 v^{a}}
-w^{a}\frac{\6}{\6 y^{a}}+f^{abc}y^{a}u^{b}\frac{\6}{\6 u^{c}}
+f^{abc}y^{a}v^{b}\frac{\6}{\6 v^{c}} \non
\+f^{abc}y^{a}w^{b}\frac{\6}{\6 w^{c}}
+\tfrac{1}{2}f^{abc}y^{a}y^{b}\frac{\6}{\6 y^{c}}
-f^{abc}w^{a}v^{b}\frac{\6}{\6 u^{c}} \ .
\eqan{EXT_DELTA_OPERATOR}

One easily verifies that $\d$ is nilpotent:

\eq
\d\d=0 \ .
\eqn{EXT_DELTA_OPERATOR_NILPOTENCY}

It should be clear at this stage that all the relevant features of
the
linear algebra \equ{ALGEBRA} are encoded in the extended operator
$\d$.
As a consequence, the analysis of possible anomalies and allowed
counterterms reduces to the study of the BRS-cohomology $\HH^{*}(\d)$
of $\d$ in the sectors of local polynomials in the fields and sources
characterized by the proper ghost charge.

\newpage
More specifically, the two distinct sectors one has to consider are:

\begin{itemize}
\item{Anomalies:}
\begin{itemize}
\item{
cohomology of $\d$ in the sector of local integrated polynomials
of dimension four and one unit of Faddeev-Popov ghost charge
\eq
\d \AA=0~~,~~\AA\not=\d \hat{\AA}~~{\rm and}~~\Phi\Pi(\AA)=1 \ ,
\eqn{EXT_DELTA_COHOMOLOGY1}
}
\end{itemize}
\item{Counterterms:}
\begin{itemize}
\item{
cohomology of $\d$ in the sector of local integrated polynomials
of dimension four and Faddeev-Popov ghost charge zero
\eq
\d \ST=0~~,~~\ST\not=\d \Delta~~{\rm and}~~\Phi\Pi(\ST)=0
\eqn{EXT_DELTA_COHOMOLOGY2}
where $\ST$ is independent of the
global ghosts ($\xi$, $\t$, u, v, w, y).
}
\end{itemize}
\end{itemize}

Now, to characterize the cohomology of $\d$ we define a filtration
operator $\NN$ as follows:
\eqa
\NN\=2\xi^{\m}\frac{\6}{\6\xi^{\m}}+2\t^{\m}\frac{\6}{\6\t^{\m}}
+2u^{a}\frac{\6}{\6 u^{a}}+2v^{a}\frac{\6}{\6 v^{a}}
+2w^{a}\frac{\6}{\6 w^{a}}+2y^{a}\frac{\6}{\6 y^{a}} \non
\+\int d^4 x \lc c^{a}\frac{\d}{\d c^{a}}+\vf^{a}\frac{\d}{\d
\vf^{a}}
+2\lp\tfrac{1}{2}B^{a}_{\m\n}\frac{\d}{\d B^{a}_{\m\n}}\rp
+2\lp\tfrac{1}{2}\chi^{a}_{\m\n}\frac{\d}{\d \chi^{a}_{\m\n}}
\rp \right. \non
\+\left.
2\eta^{a}\frac{\d}{\d\eta^{a}}+2\vfb^{a}\frac{\d}{\d\vfb^{a}}
+2b^{a}\frac{\d}{\d b^{a}}+2\cb^{a}\frac{\d}{\d\cb^{a}}
+\O^{a}_{\m}\frac{\d}{\d\O^{a}_{\m}}
+\tau^{a}_{\m}\frac{\d}{\d\tau^{a}_{\m}}\rc \ .
\eqan{FILTER_OPERATOR}
By means of a simple inspection, one notices that $\NN$ has the
structure of a counting operator which clearly induces a separation
on $\d$, namely,
\eq
\d=\sum_{n=0}^{\bar{n}} \d^{(n)}
\eqn{SEPARATION}
with,
\eq
[\NN,\d^{(n)}]=n\d^{(n)} \ .
\eqn{EIGENVALUES}
We present here the explicit expression for $\d^{(0)}$:
\eqa
\d^{(0)}\=\int d^4 x \lp\p^{a}_{\m}\frac{\d}{\d A^{a}_{\m}}
+\vf^{a}\frac{\d}{\d c^{a}}
+\tfrac{1}{2}B^{a}_{\m\n}\frac{\d}{\d \chi^{a}_{\m\n}}
+\eta^{a}\frac{\d}{\d\vfb^{a}}+b^{a}\frac{\d}{\d\cb^{a}}
+\O^{a}_{\m}\frac{\d}{\d\tau^{a}_{\m}}
+L^{a}\frac{\d}{\d D^{a}}\rp  \non
\xx\xi^{\m}\frac{\6}{\6\t^{\m}}-u^{a}\frac{\6}{\6 v^{a}}
-w^{a}\frac{\6}{\6 y^{a}}
\eqan{DELTA_ZERO}
and we remark that $\d^{(0)}$ is also nilpotent as a direct
consequence of eq.\equ{EXT_DELTA_OPERATOR_NILPOTENCY}.

One immediately notices a remarkable aspect about $\d^{(0)}$:
all fields, sources and global ghosts appear in BRS-doublets.
This implies~\cite{dix,ps,bdk} that the cohomology of $\d^{(0)}$,
acting on the unconstrained space of integrated local
field polynomials independent of the global ghosts is empty, i.e.
\eq
\HH^{*}(\d^{(0)})=\emptyset \ .
\eqn{EMPTY_COHOMOLOGY_0}
Hence, the cohomology of $\d$ vanishes as well, in view of the
fact that $\HH^{*}(\d)$ is isomorphic to a subspace of
$\HH^{*}(\d^{(0)})$. This last consequence relies on a very general
derivation developed in~\cite{dix} by using the method of spectral
sequences. Therefore we can conclude:
\eq
\HH^{*}(\d)=\emptyset \ .
\eqn{EMPTY_COHOMOLOGY_FULL}
This result implies that the Slavnov identity \equ{SLAVNOV1} is
non-anomalous
as well as the whole set of Ward identities \equ{ANTIGHOST_1},
\equ{F_CONDITION}, \equ{ANTIGHOST_2}, \equ{RIGID} and
\equ{SUSY_WARD}.
This will mean that, besides the antighost equations, also
the vector supersymmetry
Ward identity \equ{SUSY_WARD} can be employed as
a stability constraint for selecting invariant
counterterms.

Let us now turn to the computation of the invariant counterterms. The
result
\equ{EMPTY_COHOMOLOGY_FULL} already implies that they are $\delta$
variations, but this is not enough for our purpose because we have to
discard those which depend on the global ghosts. A way out is to
study the cohomology of $\BS$ and then to impose the other Ward
identities as constraints.

We first remark that the cohomology of $\BS$ in the space of local
functionals depending on all the fields (but without the global
ghosts) is empty, i.e.:
\eq
\HH^*(\BS) = \emptyset \ .
\eqn{EMPTY_COHOM_Slavnov}
The proof is immediate~\cite{wer}. We first observe that the operator
$\NN$~\equ{FILTER_OPERATOR} (with the global ghosts set to zero)
constitutes a filtration of the operator $\BS$, with $\BS^{(0)}$
coinciding with $\delta^{(0)}$~\equ{DELTA_ZERO} (global ghosts set to
zero, too). Than the proof follows from the triviality of
$\delta^{(0)}$.

Since the general counterterm $\ST$ has to obey the BRS invariance
constraint
\eq
\BS\ST = 0 ,
\eqn{BS_INVARIANCE}
we thus know that it is the $\BS$-variation of some local functional
$\Delta$ of dimension $4$ and ghost charge $-1$,
\eq
\ST=\BS\Delta \ .
\eqn{SIGMA_TILDE}

Recall now that
$\ST$ is a local integrated polynomial in the fields and sources
with dimension four and ghost charge zero. It is required
to obey the following constraints:
\eq
\ba{ll}
\WW_{\a}\ST = 0 , & \PP_{\a}\ST = 0 ,
\ea
\eqn{W_INVARIANCE}

\eq
\ba{llll}
\GG^{a}\ST = 0 , & \FF^{a}\ST = 0 , &
\GG^{\prime a}\ST = 0 , & \RR^{a}\ST = 0 ,
\ea
\eqn{G_INVARIANCE}

\eq
\ba{llll}
\dfrac{\d\ST}{\d\eta^{a}} = 0\ ,
               & \dfrac{\d\ST}{\d b^{a}} = 0\ ,&
\dfrac{\d\ST}{\d\vfb^{a}}-\6^{\m}\dfrac{\d\ST}{\d\tau^{a\m}} = 0\ ,
  \dfrac{\d\ST}{\d\cb^{a}}-\6^{\m}\dfrac{\d\ST}{\d\O^{a\m}} = 0\ .
\ea
\eqn{ETA_INVARIANCE}

Let us here make two comments:
Firstly, one has to notice that all possible quantum corrections to
$\S$
are given by BRS-coboundaries: this assures that the BRS-triviality
of the classical action is preserved at the quantum level.
Secondly, since there are no physical parameters in the model, one is
essentially interested in the study of anomalous dimensions, these
latters being related to BRS-trivial counterterms.

$\ST$ obeying the constraints above can we choose $\Delta$ to obey
them, too ? It is clear that, since the set of constraints above is
stable under the action of $\BS$ (see~\equ{ALGEBRA}), $\ST$ will obey
these constraints if $\Delta$ does. Although $\Delta$ does not have
to fulfill them it may be chosen to do so, except for the constraint
given by the third eq. of~\equ{G_INVARIANCE}.
This is shown in Appendix
A for the conditions given by the fourth eq. of~\equ{G_INVARIANCE}
and by eqs.~\equ{ETA_INVARIANCE} and in Appendix B for the first two
of~\equ{G_INVARIANCE}.
To summarize we can write
\eq
\GG^a\Delta= \FF^a\Delta=
\RR^{a}\Delta =
\frac{\d\Delta}{\d\eta^{a}} =
\frac{\d\Delta}{\d b^{a}} =
\frac{\d\Delta}{\d\vfb^{a}}-\6^{\m}\frac{\d\Delta}{\d\tau^{a\m}} =
\frac{\d\Delta}{\d\cb^{a}}-\6^{\m}\frac{\d\Delta}{\d\O^{a\m}} = 0\ .
\eqn{SUMMARY}

This gives rise to a rather small set of independent counterterms
$\ST$. This set is further reduced by applying the last
constraints,
namely the third of eqs.~\equ{G_INVARIANCE} and, last but not least,
the supersymmetry constraint~\equ{W_INVARIANCE}
(translation invariance
is obvious).

Finally one is able to
express the most general local conterterm $\ST$ as:
\eq
\ST=\ST^{(2)}+\ST^{(3)}
\eqn{GENERAL_COUNTERTERM}
with,
\eqa
\ST^{(2)}\=\BS\int d^4 x \{ a_{1}\lc\lp\O^{a\m}-\6^{\m}\cb^{a}\rp
A^{a}_{\m}
+\lp\tau^{a\m}-\6^{\m}\vfb^{a}\rp \psi^{a}_{\m}\rc  \non
&&~~~~~+a_{2}\lp\tau^{a\m}-\6^{\m}\vfb^{a}\rp\6_{\m}c^{a}
+a_{3}\chi^{a\m\n}\6_{\m}A^{a}_{\n} \}
\eqan{ST2}
and
\eqa
\ST^{(3)}\=\BS\int d^4 x\lac b
f^{abc}\chi^{a\m\n}A^{b}_{\m}A^{c}_{\n}\rac
\eqan{ST3}
where $a_1$, $a_2$, $a_3$ and $b$ are the arbitrary coefficients
of the four possible counterterms.
We emphasize that no other combination of fields and sources would be
compatible with
eqs.\equ{BS_INVARIANCE},\equ{W_INVARIANCE}-\equ{ETA_INVARIANCE}.

An interesting result of the present study is that also with the
partial-derivative Landau type gauge used here \equ{TOP_GAUGE},
the counterterm
\eq
\int d^4 x (F_{\m\n}^{+})^{2}
\eqn{COUNTER}
does not show up. This agrees with~\cite{wer} and with a previous
1-loop computation carried out in~\cite{bbt,bbrt}.

\section{Concluding Remarks}

In this paper we have investigated the issue of the renormalization
of
Witten's topological Yang-Mills field theory in the presence of
several invariances. In particular, it has been made direct use of a
vector supersymmetric Ward identity which, besides the Slavnov and
the other identities, has been shown to be free of anomalies. As a
second step, one has proceeded to the calculation of the most general
local counterterm of the model \equ{GENERAL_COUNTERTERM},
compatible with those symmetry constraints.

The topological vector supersymmetry deminishes the number of
potential UV divergences much more drastically than in~\cite{wer}
where the constraint of the vector symmetry could not be imposed.
Nevertheless, one ends up with a set of BRS-trivial
monomials related to the anomalous dimensions of the model, whose
coefficients may be determined by means of Feynman diagrams. The
conclusion is that the cohomological nature of the model remains
unaltered at the quantum level.

\section{Acknowledgements}

We would like to thank Silvio P. Sorella for many useful comments and
suggestions. M.W.O. is grateful to the ``Fonds zur F\"orderung der
Wissenschaftlichen Forschung'' for a Lise Meitner grant.


\section*{Appendix A}

\setcounter{equation}{0}
\renewcommand{\theequation}{A.\arabic{equation}}

In this appendix we show that there is no loss of generality if one
assumes that the local polynomial $\Delta$, defined in
eq.\equ{SIGMA_TILDE},
\eq
\ST=\BS\Delta \ ,
\eqn{3.13}
equally obeys the constraints given by the fourth
eq. of
\equ{G_INVARIANCE} and eqs. \equ{ETA_INVARIANCE} which are
originally
verified by the counterterm $\ST$.
In other words $\Delta$ can be taken as a rigid invariant object
(i.e. a trace in the adjoint representation of the gauge group
$\bf{G}$)
satisfying the set of gauge fixing and ghost conditions obtained
previously.

We begin by analyzing the rigid gauge invariance $\RR^a$. As a
consequence of
its $\RR^a$-invariance $\ST$ is also invariant under the action of
the
quadratic Casimir operator, i.e.
\eq
\RR^2 \ST \equiv \RR^a \RR^a \ST = 0 \ .
\eqn{R_NILPOTENCY}
Recalling the commutation relation
\eq
\lc\RR^a,\BS\rc=0 \ ,
\eqn{R_BS_COMMUTATOR}
one may set:
\eq
\BS \RR^2 \Delta = 0 \ .
\eqn{BSRR_DELTA}
The local polynomial $\Delta$ can be
split up in two parts:
\eq
\Delta = \Delta^\natural + \Delta^\flat \ ,
\eqn{DELTA1}
where $\Delta^\natural$ is invariant and $\Delta^\flat$ is
non-invariant under rigid transformations. The non-invariant part
decomposes in terms of the eigenvalues of $\RR^2$:
\eq
\Delta^\flat=\sum_{l=1}^{\bar{l}}\Delta_{l}
\eqn{DELTA2}
with
\eq
\RR^2\Delta_{l}=c_{l}\Delta_{l} \ .
\eqn{RR_EVE}
The eigenvalues $c_{l}$ are strictly positive constants, the
generators being represented by hermitian matrices. Equation
\equ{BSRR_DELTA}
can then be rewritten as follows:
\eq
\sum_{l=1}^{\bar{l}}c_{l}\BS\Delta_{l}=0 \ .
\eqn{DELTA3}
Since $\BS$ commutes with the rigid transformations and the
$\Delta_l$ are independent of each other, the unique solution
to \equ{DELTA3} is
\eq
\BS\Delta_{l}=0\ ,~~~~~~~\forall l \ge 1\ ,
\eqn{BSD1}
so that
\eq
\BS\Delta^\flat=0 \ .
\eqn{BSD2}
Hence, the non-invariant part $\Delta^\flat$ is $\BS$-invariant,
giving
no contribution to $\ST$.

We now discuss the extension to $\Delta$, of the validity of
the gauge fixing and ghost conditions \equ{ETA_INVARIANCE}.
The formers read as:
\eq
\frac{\d\ST}{\d b^a}=\frac{\d\ST}{\d \eta^a}=0 \ ,
\eqn{GAUGE_DELTA}
and, by using the redefinitions
\eqa
\label{GHOST_COMBINATION1}
\hat{\O}^{a}_{\m}\=\O^{a}_{\m}-\6_{\m}\cb^{a} \ , \\
\hat{\tau}^{a}_{\m}\=\tau^{a}_{\m}-\6_{\m}\vfb^{a} \ ,
\eqan{GHOST_COMBINATION2}
one may rewrite the ghost equations as
\eq
\frac{\d\ST}{\d \cb^a}=\frac{\d\ST}{\d \vfb^a}=0.
\eqn{GHOST_DELTA}
Now, we introduce the filtration $F$:
\eq
F=\int d^4 x \lp b^a\frac{\d}{\d b^a}+\cb^a\frac{\d}{\d \cb^a}
+\eta^a\frac{\d}{\d \eta^a}+\vfb^a\frac{\d}{\d \vfb^a}\rp =
\lac \BS , \cb^a \frac{\d}{\d b^a} + \vfb^a \frac{\d}{\d \eta^a}
\rac
\eqn{FILTER1}
and we notice that it commutes with the Slavnov operator,
\eq
\lc F,\BS \rc =0 \ .
\eqn{FILTER2}
One has to observe that $F$ will induce a separation on $\Delta$:
\eq
\Delta=\sum_{m=0}^{\bar{m}}\Delta^{(m)}
\eqn{DELTA_F1}
with
\eq
F\Delta^{(m)}=m\Delta^{(m)}.
\eqn{DELTA_F2}
{}From eq.\equ{3.13} one can write the counterterm as:
\eq
\ST=\BS\lp\Delta^{(0)}+\sum_{m=1}^{\bar{m}}\Delta^{(m)}\rp
\eqn{SIGMA_T2}
and, as a consequence of \equ{GAUGE_DELTA} and \equ{GHOST_DELTA}, we
have:
\eq
F\ST=0 \ .
\eqn{FST}
By combining eqs.\equ{SIGMA_T2} and \equ{FST}, we get:
\eq
\sum_{m=1}^{\bar{m}}m\BS\Delta^{(m)}=0 \ .
\eqn{CONDITION}
Since the sectors $\Delta^{(m)}$ remain independent under the
action of $\BS$, there is only one possible solution to
\equ{CONDITION}, namely
\eq
\BS\Delta^{(m)}=0~~~~~~~\forall m\ge 1 \ .
\eqn{BSD_M}
Therefore, we see that all terms $\Delta^{(m)}$ with weights
greater than zero play no role in the construction of $\ST$.

To summarize, we have obtained the following result :
\eq
\ST = \BS\Delta\ ,
\eqn{Result}
with $\Delta$ obeying the constraints given by the rigid invariance,
the gauge condition and the ghost equation.

\section*{Appendix B}

\setcounter{equation}{0}
\renewcommand{\theequation}{B.\arabic{equation}}

This appendix contains a detailed description of the use of the
antighost equation and of the $\FF$-Ward identity as constraints on
the evaluation of $\ST$ (see the first two eqs.
of~\equ{G_INVARIANCE}).

The study of this constraints is greatly simplified if one
performs the following field redefinitions
\footnote{Here $\O_{\m}$ and $\tau_{\m}$ stand for the redefined
variables of
\equ{GHOST_COMBINATION1} and \equ{GHOST_COMBINATION2}.}:
\eqa
\hat{\psi}_{\m}^{a}\=\psi_{\m}^{a}+f^{abc}c^{b}A_{\m}^{c} \ , \\
{\hat\O}_{\m}^{a}\=\O_{\m}^{a}
  +f^{abc}c^{b}\tau_{\m}^{c} \ , \\
\hat{\vf}^{a}\=\vf^{a}+\tfrac{1}{2}f^{abc}c^{b}c^{c} \ , \\
\hat{L}^{a}\=L^{a}+f^{abc}c^{b}D^{c} \ .
\eqan{REDEFINITIONS}
Thereby, in terms of these shifted variables, the first two
conditions
\equ{G_INVARIANCE} are replaced by:
\eq
\int d^4 x \frac{\d\ST}{\d\hat{\vf}^a}
=\int d^4 x \frac{\d\ST}{\d c^a}=0 \ .
\eqn{GHOST_EQU_3}
Now, in order to avoid ambiguities, it
behooves
us to translate the
conditions \equ{GHOST_EQU_3} above to a non-integrated level.
To do this, we properly substitute functional derivatives by ordinary
ones, introducing a prescription in order to handle with $\ST$ and
$\D$ at the level of their integrands.

The general structure of the local counterterm $\ST$ to be understood
here is:
\eq
\ST=\sum_{k=1}^{\bar{k}}\int d^{4}x M_{k}
(c,\6 c,...;\hat{\vf},\6 \hat{\vf},...)X_{k}
\eqn{GENERAL_ST}
where $M_k$ is a monomial depending on $c^a$, $\hat{\vf}^a$ and their
space-time derivatives; $X_k$ is another monomial built up from
all other fields and their derivatives.
The associated integrand $\tilde{\o}$ of $\ST$ is then defined to be:
\eq
\tilde{\o}=\sum^{\bar{k}}_{k=1}\tilde{\o}_k
\eqn{OMEGA_TILDE1}
with,
\eq
\tilde{\o}_k=\lac
\begin{array}{ccc}
M_{k} X_{k}  & $if$ & X_k\not=\6 Y_k \\
{}~&~&~\\
-\6 M_k Y_k  & $if$ & X_k=\6 Y_k
\end{array}\right.
\eqn{OMEGA_KAPPA}
and, repeatedly, if $X_k=\6\6 Y_k$ then, $\tilde{\o}_k=\6\6 M_k Y_k$,
and so on. With this
prescription
in mind, one is allowed to rewrite
expression \equ{3.13} in a unique way:
\eq
\tilde{\o}=\BS\o \ ,
\eqn{OMEGA_TILDE2}
$\o$ being the integrand of $\D$, defined according to the same rule.
That prescription \equ{OMEGA_KAPPA} is unique, can be easily seen
from the fact that:
\eqa
\ST=\int d^{4}x \tilde{\o} = 0
{}~~~~~\Rightarrow~~~~~\tilde{\o}=0 \ . \nonumber
\eqan{FACT}
Indeed, one has in general the following:
\eqa
\ST=0~~~~~\Rightarrow~~~~~\ST
\=\int d^4 x \6\lp\sum^{\bar{k}}_{k=1}M_k(c,...;\hat{\vf},...)X_k\rp
\non
\=\int d^4 x \lp\sum^{\bar{k}}_{k=1}\6 M_k X_k
+\sum^{\bar{k}}_{k=1} M_k \6 X_k\rp \nonumber
\eqan{GENERAL_SIGMA_TILDE}
where $X_k$ may be a total space-time derivative or not. If yes,
i.e. if
\eqa
X_k=\6 Y_k  \nonumber ,
\eqan{TOTAL_DERIVATIVE}
then, with \equ{OMEGA_KAPPA}
\eqa
\tilde{\o}_k=-\6\6 M_k Y_k+\6\6 M_k Y_k=0 \ . \nonumber
\eqan{OMEGA_TILDE_KAPPA1}
On the other hand, if $X_k$ is non-differentiated, one has again:
\eqa
\tilde{\o}_k=\6 M_k X_k-\6 M_k X_k=0 \ . \nonumber
\eqan{OMEGA_TILDE_KAPPA2}
At the level of the integrands, the constraints
\equ{GHOST_EQU_3} read
\eq
\frac{\6\tilde{\o}}{\6 \hat{\vf}^a}=\frac{\6\tilde{\o}}{\6 c^a}=0\ .
\eqn{WT_CONSTRAINTS}
We now define the filtration,
\eq
\stackrel{\circ}{F}~=c^a\frac{\6}{\6 c^a}
+\hat{\vf}^a\frac{\6}{\6 \hat{\vf}^a}
\eqn{FILTERING}
which commutes with $\BS$:
\eq
\lc\stackrel{\circ}{F},\BS\rc=0 \ .
\eqn{FILTERING_COMMUTATOR}
{}From \equ{WT_CONSTRAINTS} follows
\eq
\stackrel{\circ}{F}\tilde{\o}=0 \ .
\eqn{FWT}
Since the filtration induces a separation on $\o$,
one can write eq.\equ{OMEGA_TILDE2} as:
\eq
\tilde{\o}=\BS\lp\o^{(0)}+\sum_{p=1}^{\bar{p}}\o^{(p)}\rp \ .
\eqn{WT1}
Applying the filtration on both sides of \equ{WT1}
implies
\eq
\sum_{p=1}^{\bar{p}} p\BS\o^{(p)}=0 \ ,
\eqn{WT2}
with one single solution,
\eq
\BS\o^{(p)}=0~~~~~~~\forall p \ge 1 \ .
\eqn{SOLUTION}
We conclude then that
\eq
\tilde{\o}=\BS\o^{(0)} \ .
\eqn{SOLUTION_WT}
Hence, coming back to the integrated level, one can disregard
the contributions
stemming from monomials in $\D$ which depend on $c^a$ and
$\hat{\vf}^a$. Thus
\eq
\ST=\BS\D^{(0)} \ ,
\eqn{FINAL}
$\Delta^{(0)}$ being independent of $c^a$ and ${\hat\varphi}^a$
obeys the constraints \equ{GHOST_EQU_3}.

\end{document}